\documentclass[11pt,a4paper]{article}
\usepackage{amsmath,amssymb}
\usepackage{graphics}
\usepackage{epsfig}
\usepackage{graphicx}
\newcommand{\be}{\begin{equation}}
\newcommand{\ee}{\end{equation}}
\newcommand{\bea}{\begin{eqnarray}}
\newcommand{\eea}{\end{eqnarray}}
\newcommand{\beas}{\begin{eqnarray*}}
\newcommand{\eeas}{\end{eqnarray*}}
\newcommand{\ba}{\begin{array}}
\newcommand{\ea}{\end{array}}
\newcommand{\nn}{\nonumber}

\newcommand{\bt}{\begin{table}}

\newcommand{{\vsi}}{\varsigma}

\newcommand{\al}{\alpha}
\newcommand{\ga}{\gamma}

\newcommand{\de}{\delta}
\newcommand{\De}{\Delta}

\newcommand{\ka}{{\kappa}}
\newcommand{\la}{\lambda}
\newcommand{\La}{\Lambda}
\newcommand{\na}{\nabla}

\newcommand{\si}{\sigma}

\begin{document}
\title{Multiscalar-metric gravity: cosmological constant 
screening and emergence of massive-graviton 
dark components of the Universe}
\author{Yury F.~Pirogov\footnote{E-mail: pirogov@ihep.ru}
\\
{\small
Theory Division, Logunov Institute for High Energy Physics of 
NRC  ``Kurchatov Institute'',
}\\
{\small
\em 
Protvino, 142281 Moscow Region, Russia
}\\ 
%\small{\em E-mail:  pirogov@ihep.ru}
}
\date{}
\maketitle

\begin{abstract}
\noindent
In the multiscalar-metric  frameworks, 
the  issues of the vacuum energy/cosmological constant (CC)  screening due to   
the  Weyl-scale enhancement of the Diff gauge symmetry, along with  emergence of the massive 
dark gravity components through  the gravitational  Higgs mechanism are considered. 
A   generic dark gravity model is developed, with  
the two extreme versions of the model of particular interest based on 
General Relativity (GR) and its classically equivalent Weyl transverse alternative  
being compared and argued  to be, generally, inequivalent. 
The so constructed spontaneously broken Weyl Transverse Relativity (WTR)
is proposed as a viable beyond-GR effective
field theory of gravity, with screening of the  Lagrangian CC, 
superseded by the induced one, and   emergence of the
massive tensor and scalar gravitons as the dark gravity components.
A basic concept with  the spontaneously broken  Diff gauge symmetry/relativity 
-- in particular,  WTR vs.\ GR -- as a principle source of the emergent 
dark gravity components of the Universe is put forward.\\


 %{\bf PACS:} {{04.50.Kd} {Modified theories of gravity}; % \and 
%{95.35.+d} { %\and
%{95.36.+x}  { Dark energy}
%} % end of PACS codes
%} %end of abstract

\end{abstract}

\section{Introduction}

The  issue of the vacuum energy/cosmological constant (CC) in General Relativity (GR) 
due to the so-called zero-point quantum fluctuations is well-known 
already for a long time.\footnote{For an original treatment 
of the vacuum energy/CC problem, see,~\cite{Zeld}.}
But only recently, in the wake of  the appearance of the convincing observational evidences for (an effective) CC responsible  
for  the late-time accelerated expansion of the Universe, as expressed 
by the cosmological  Standard Model~(SM), or, otherwise, the LCDM model,
the   problem  of CC (or, more generally, the vacuum energy) 
seems to become the most crucial one in the realm of 
the contemporary fundamental physics.\footnote{For the observational status of CC, see, 
e.g.,~\cite{Carr}. For the modern view of the vacuum energy/CC problem, see, e.g.,~\cite{Wein, Mart}.}  
Though not meaning any explicit discrepancy with observations, this  problem 
signifies  nevertheless at least a tension  between 
the two  present-day  basic physical theories: 
the theory of gravity -- GR   
and the theory of matter and its interactions --- the quantum field theory (QFT) ---  
expressed more particularly by the particle SM. 
Namely, in the  effective field theory (EFT) framework  
the CC problem is, in fact, at least threefold:\footnote{For the importance of  the EFT 
approach to the CC problem, cf, e.g.,~\cite{Burg}.}  

(\/{\em I}\/) {\em The (un)naturalness problem\/}: 
What forbids  the huge  Lagrangian CC, expected in the EFT framework, 
from the explicit manifestations, with a non-zero tiny observational   CC being (technically) ``unnatural''?

(\/{\em II}\/) {\em The  coincidence problem\/}: 
What defines such a  tiny value of the observational CC, so that the latter manifests itself 
in the accelerated  expansion of the Universe  just ``lately'' on the cosmological time scale? 

(\/{\em III}\/) {\em The quantum stability problem\/}: 
What prevents the tiny observational CC from being drastically renormalized  by the radiative  corrections,
a priori expected in  the EFT framework?

Solving these and related problems presents 
a challenge to  the modern fundamental physics and could ultimately imply 
either a revision of QFT, or a GR modification, with a variety  of the latter  
ones  proposed to this end up to now.\footnote{For the modifications of  gravity, see, e.g.,~\cite{Capo}--\cite{Noj}   
and, in particular,~\cite{Rubak, de Rham} for the massive gravity.} 
More particularly, to solve the CC problem the modified gravity should,  conceivably, be treated   as EFT 
valid at the relatively low energy scales (compared to the fundamental one given by the Planck mass).
At that, GR as EFT is well-known, first,  to be 
constructed from a symmetric second-rank tensor field -- the metric $g_{\mu\nu}$, 
with its determinant $g$
defining a four-volume element $\sqrt{-g}d^4 x$ (in the spacetime dimension  $n=4$), 
with  the spacetime measure $\sqrt{-g}$,
and, second, to be  based on  the gauge symmetry of the (full) diffeomorphisms (Diff's) incorporating, in particular, 
the longitudinal ones. 
The fact that the spacetime measure $\sqrt{-g}$ transforms under  the longitudinal diffeomorphisms 
proves to be  in GR (and its  siblings) an ultimate reason of the emergent CC problem. 
By this token, we take the issue of $g$ as a guide for choosing
a proper route towards  solving the CC problems, with some principle steps  
leading to this goal shortly indicated below.

(\/{\em i}\/) {\em Unimodular relativity.} 
First of all,  long before the CC problem had become so acute, 
for solving  such a  problem  there was proposed  in~\cite{And} (with
the numerous subsequent elaborations) to  substitute  GR 
based on  the full Diff symmetry  by  Unimodular Relativity (UR), known also as Unimodular Gravity, 
based on the  transverse diffeomorphism (TDiff $\subset$  Diff) symmetry.\footnote{For the 
relevance of the TDiff gauge symmetry as a substitute of the full Diff one for the consistent description of the 
massless tensor graviton, see~\cite{Ng}.}
By construction, instead of the conventional ten-component  
(at $n=4$) metric  $g_{\mu\nu} $, 
UR/TDiff uses  a reduced nine-component one $\hat g_{\mu\nu} $, with the  determinant $\hat g={\ka}$  
given by a fixed absolute/non-dynamical scalar density ${{\ka}}$ 
(for simplicity, say,  ${\ka}=-1$).\footnote{The notation in Introduction   is 
in accord with what follows.}  
The latter defines  simultaneously the spacetime volume element $\sqrt{-\ka}\, d^4 x$.
The Lagrangian CC within UR proves to be  irrelevant being superseded by 
an integration constant
having yet no clear-cut  physical meaning. 
Due to this, UR remains, in a sense,  dynamically incomplete. 
Nevertheless, solving the first part of the CC problem, UR  definitely presents a  step in the right direction,  
though  hardly sufficient to completely solve the problem, and implies further elaboration.

(\/{\em ii}\/) {\em Weyl transverse relativity.} 
In the latter respect, in~\cite{Iza}  there  was proposed   
(cf.\ also~\cite{Alv1}) a ten-component modification of UR constructed from 
a relative tensor  $ \hat  g_{\mu\nu} \equiv  g_{\mu\nu}/(-g)^{1/4}$ (at $n=4$), 
with a fixed determinant (put for definiteness  $\hat g=-1$),
the former being liable to be  converted into the true tensor through multiplication by $(-{\ka})^{1/4}$.
By construction, the respective  EFT of gravity --- to be called  Weyl Transverse Relativity  (WTR)  ---  
satisfies a gauge symmetry, WTDiff, consisting of 
TDiff enhanced by the Weyl scale transformations, and 
provides ultimately  an improved, compared to  UR, 
solution of the CC problem.\footnote{For a development of WTR,  known also as  
Weyl Transverse Gravity, cf., e.g.,~\cite{Alv2, Oda1}.} 
Likewise UR/TDiff,   its modification WTR/WTDiff  may  naturally  give a justification  
of the disappearance of the  (huge) Lagrangian CC.
At the classical level (under the covariant conservation of the energy-momentum tensor) 
WTR/WTDiff proves to be equivalent to GR/Diff,   
with an arbitrary induced CC to be treated as a global degree of freedom.
Moreover at the quantum level, WTR/WTDiff could ensure 
the quantum stability of the residual classical CC against 
the  radiative corrections~\cite{Carb1, Alv3}.
Besides, on the field-theoretic side, 
WTR may be obtained as a viable alternative to GR proceeding   
in a  self-consistent manner, in the complete analogy  to GR, 
from  a theory of the  precisely two-d.o.f.\  self-interacting  
massless graviton~\cite{Carb2}.\footnote{This fact 
alone could, conceivably,  justify WTR (and its siblings) 
to be further explored at least on par with GR (and its siblings).}   
Altogether, solving  (at least partially) the first and the third parts of the CC problem  
and possessing a solid field-theoretic status  WTR/WTDiff could thus 
constitute a viable alternative to GR/Diff,
realizing one more  step  in the  right direction, 
with the next one(s) still needed.
More particularly, within  WTR  per se  (likewise UR) there appears no indication  
on a infrared  mass scale for the  tiny residual CC, 
with the second part of the CC problem remaining to be resolved,~too.

(\/{\em iii}\/) {\em Higgs mechanism for gravity.} 
The looked-for infrared  mass scale could naturally be provided  
by a massive  graviton, 
the latter  serving as the putative dark energy (DE) mimicking  the observed CC. 
At that, the original tensor-graviton mass term due to Fierz and Pauli~\cite{Fierz} explicitly violates 
the Diff gauge symmetry, and was later shown to result in the massive-massless perturbative discontinuity~\cite{van Dam, Zakh},
to remedy which there was proposed~\cite{Vain} a nonperturbative mechanism for recovering the transition.
It was also found~\cite{Deser} that such an explicit Diff gauge symmetry violation 
results in the appearance of a ghost, with  a violation of unitarity.
Thus, with the massless GR  treated as a Diff gauge theory, the consistent ghost-free theory of massive gravity would, 
conceivably, imply not the explicit violation but the spontaneous breaking  of such a gauge symmetry  
by a counterpart of the Higgs mechanism for gravity. 
An  attempt to approach  such a  goal, modifying GR 
through the Higgs mechanism in terms  of a  quartet  of the scalar fields, 
was undertaken  in~\cite{tHooft} 
(see also \cite{Kaku1, Demir}),
in the wake of which there was  proposed in~\cite{Cham1}--\cite{Oda3}
a simple solution to the problem elaborated further in~\cite{Cham2, Cham3}.
In this  respect, GR augmented by  the Higgs mechanism for gravity  
could naturally   provide a required  scale  for the  late exponential expansion of the Universe,   
associating such a scale   with the tiny tensor-graviton mass. 
The latter, on the other hand, should naturally be treated as a parameter of the theory (given still ad hoc)  
likewise any other fundamental parameter of Nature. 
Out of the four (at $n=4$) scalar fields  used in such a Higgs mechanism within  GR, 
the three components 
can, at a proper choice of the (quasi-)Higgs potential,  be absorbed as the additional components of the massive tensor graviton.
At that, one more component may be  stated  to  become a ghost independently  of the potential, 
though  in a sufficiently high  order of  the perturbation theory~\cite{Cham3}.
Besides,  in the massive-modified  GR there still remains unclear the reason of absence of the (huge) Lagrangian CC,
as well as the radiative stability  of the Higgs mechanism (in the absence of a protecting symmetry).  
So, the Higgs mechanism in GR is  on its own hardly sufficient  to  exhaust  the CC problem,~too.

(\/{\em iv}\/) {\em Quartet-metric gravity.} 
Altogether,  merging the above mentioned directions 
of the gravity modification  to  reconcile the various parts of the CC problem, 
with   the two principle ingredients of the properly modified gravity ---  
an effective metric $\hat g_{\mu\nu}$, $\hat g={\ka}$,  
and a (quasi-)Higgs field  --- 
to stem from the same source  --- the scalar quartet --- may be obligatory.\footnote{This, in turn, ensures
one more important ingredient for solving the CC problem -- the dynamical non-gravitating 
spacetime measure  $\sqrt{-{\ka}}$.
For the relevance   of such a measure 
built of the so-called ``measure fields''  within  the modified GR 
still with the conventional metric $g_{\mu\nu}$, cf., e.g.,~\cite {Guendel1, Guendel2}.}  
A new concept should, conceivably, be invoked to this end, 
with a change of the spacetime paradigm~\cite{Pir1, Pir2}.
According to the latter,  among  the arbitrary kinematic observer's coordinates there exist 
some  distinct dynamical ones, associated ultimately with the vacuum and 
given by a quartet  (at $n=4$) of the peculiar (invertible) scalar fields.
Such a  scalar quartet is to be considered on par with the metric 
as a basic field variable to construct EFT of the  so-called  quartet-metric gravity.
Introducing in such a framework, on the one hand, a proper effective metric, without adding 
the (quasi-)Higgs potential for the massive tensor graviton,  one  can 
eliminate the (huge) Lagrangian CC as in WTR.  
On the other hand,  introducing the proper  (quasi-)Higgs potential 
without modifying  metric, one can produce, as in GR,  
the massive tensor graviton serving, conceivably, as DE.
Superimposing these two routes of the gravity modification within the quartet-modified WTR 
may, conceivably, solve both the problem  of screening of the (huge) Lagrangian CC and 
emergence of the (tiny) observable  one supplemented by the associated dark gravity components 
-- the massive tensor and scalar gravitons.
This is  a goal of the present paper.

In section~2, the basics of the quartet-metric (dark) gravity are exposed.
In section~3,  a sufficiently general prototype dark gravity model is constructed   
to realize  the concept of the spontaneously broken Diff gauge symmetry/relativity 
as a source of the emergent dark gravity components. 
The two extreme  versions 
of such a  generic model of the particular interest corresponding, respectively, to the spontaneous breaking  of  GR 
vs.\ its classically equivalent alternative --- WTR --- are worked out in sections~4 and 5.
These  versions are compared in the context 
of the CC screening and  emergence of the  dark gravity components through the gravitational Higgs mechanism, 
and are argued to be, generally, non-equivalent.
The consistency  of the so constructed   spontaneously broken WTR  as  
a theory  of the massive tensor and scalar gravitons is demonstrated in the weak-field limit.
In Conclusion, the viability of 
the spontaneously broken WTR 
as a  beyond-GR EFT of gravity 
solving the CC problem  through  screening  the Lagrangian CC, superseding it  by  
the induced  one and supplementing by the massive tensor and scalar 
gravitons  as the dark gravity components is argued.
The importance of further developing the basic concept of  the spontaneously broken Diff gauge symmetry/relativity 
--- in particular,  WTR vs.\ GR --- 
as a principle reason of  emergence of the  dark gravity components of the Universe is stressed.\footnote{Though 
adhering primarily to the spacetime dimension $n=4$,
not to be formally confined  by this value,
the term ``scalar quartet''  valid at $n=4$  is substituted in what follows
by  a more general one --- the ``multiscalar'' --- and, accordingly, ``quartet-metric'' by ``multiscalar-metric'',~etc.}

\section{Multiscalar-metric  dark gravity: basics}

\subsection{Multiscalar-extended  field set}

Under the term ``dark gravity'' there will be   understood  a merger 
of the (conventional) gravity with  the emergent dark gravitational components.
The EFT of the multiscalar-metric dark  gravity may 
basically be  defined through  the generally covariant  action functional\footnote{Here we start directly from the EFT level. 
For the justification of EFT of the multiscalar-metric gravity as an affine-Goldstone nonlinear model, see~\cite{Pir2'}.}
\be\label{S}
S[g_{\mu\nu},  {{X}}^a]  = \int {\cal L}(g_{\mu\nu}, {{X}}^a) \, d^n x 
\ee
in terms of a Lagrangian density $\cal L$ dependent on a tensor field --- the basic metric ---  $g_{\mu\nu}$ 
(possessing by an inverse $g^{-1\mu\nu}$) extended through  a set of the scalar fields   
$X^a$, with $a,b,\dots =0,1, \dots, n-1$ being the  global 
Lorentz indices and $n$ a spacetime dimension.
The scalar multiplet  $X^a=X^a(x)$ --- the multiscalar ---
is peculiar by the fact of being (in a patch-wise manner) 
invertible,  $x^\mu =x^\mu(X)$.
This allows to use it as some distinct dynamical  spacetime 
coordinates $X^\al\equiv \de^\al_a {X}^a(x)$ 
among the arbitrary kinematic ones $x^\mu$,
as well as to introduce  the respective dynamical  frame
$X^\al_\mu\equiv \partial_\mu X^\al$, $\det (X^\al_\mu)\neq 0$, 
possessing by an  inverse~$ X^{-1}{}_\al^\mu \equiv X_\al^\mu $. Stress that  
such a  multiscalar is not an auxiliary (non-dynamical) variable introduced just to restore  
the general covariance, but a faithful dynamical variable on par with the basic metric. 
Ultimately, the dynamical coordinates
$X^\al$ may be  treated as  those associated  with the (dynamical) vacuum.
The action $S$ is moreover postulated to be invariant under the reparametrizations of ${X}^a$ 
given by the Lorentz transformations and the finite constant shifts, $X^a\to X^a+C^a$.
By this token, ${X}^a$ should, in fact, enter through 
a  derivative term ${\ka}_{\mu\nu}$  constructed as 
\be\label{om}
{\ka}_{\mu\nu}
\equiv  \partial_\mu  {X}^a  \partial_\nu {X}^b \eta_{ab},
\ee
with $\eta_{ab}$ being  the Lorentz-invariant Minkowski symbol.
By default, ${\ka}_{\mu\nu}$, serving as a  (quasi-)metric,  possesses by  an inverse  ${\ka}^{-1\mu\nu}$ 
(at least  patch-wise with the proper matching), 
with 
\be
{\ka}\equiv    \det({\ka}_{\mu\nu})  =(\det ( \partial_\mu  {X}^a ))^2 \det(\eta_{ab}) <0
\ee
to ensure the  non-degeneracy  and invertibility of such a (quasi-)metric.\footnote{To this end, we assume 
the  (patch-wise) Lorentz-preservation in the field space of $X^a$, 
up to the physically equivalent affine redefinitions of the  fields $X^a$ and the metric $\eta_{ab}$.} 
In terms of the independent field variables, the multiscalar-metric action, in fact, looks like 
\be\label{S1}
S[g_{\mu\nu},  {{X}}^a]  = \int {\cal L}(g_{\mu\nu}, \ka_{\mu\nu}) \, d^n x .
\ee
Further, in terms of ${\ka} $ and $g=\det(g_{\mu\nu})<0 $
one can define an effective  scalar field 
\be\label{scal}
\si=\ln (\sqrt{-g}/\sqrt{-{\ka}}),
\ee
to be called the scalar graviton.
By this token, in terms  of the fields $ g_{\mu\nu}$ and $ {\ka}_{\mu\nu}$ 
one can  introduce the effective ones  
as  a conformal metric $\bar g_{\mu\nu}$ and a tensor field
$\bar {{\ae}}^\mu_\nu$ 
given by  a correlator (partial contraction) of the two metrics as follows:
\be\label{conform}
\bar g_{\mu\nu}\equiv e^{{\bar {w}}}  g_{\mu\nu}, \quad 
\bar {{\ae}}^\mu_\nu \equiv \bar g^{\mu\la} {\ka}_{\la\nu},
\ee
where  
${\bar {w}}={\bar {w}}(\si)$ is a gravitational scale factor,\footnote{The dependence  $\bar w(\si)$  presents the  simplest case. 
More generally, one could envisage  within EFT the dependence of the effective metric scale $\bar w$ 
on the matter fields, environment, etc, with the ultraviolet and infrared behavior of the effective metric being a priori quite different.} 
 $\bar g^{\mu\nu}\equiv \bar g^{-1\mu\nu}
=e^{-{\bar {w}}} g^{-1\mu\nu}$  is an  inverse  of $\bar g_{\mu\nu}$ and 
$\det(\bar {{\ae}}^\mu_\nu)={\ka} /\bar g=e^{-2(2{\bar {w}} +\si)}$, 
$\bar g\equiv \det(\bar g_{\mu\nu})<0$.\footnote{By default, the spacetime 
indices are now lowered (raised) through the effective metric $\bar g_{\mu\nu}$ ($\bar g^{\mu\nu}$)
if not stated explicitly otherwise.}
The tensor field $\bar {{\ae}}^\mu_\nu$ may be  treated as a realization of  the (a priori  loose) DE.
By means  of these effective fields,  merging  gravity and its dark components,  the effective action  
of the purely multiscalar-metric dark gravity may be presented quite generally as 
\be\label{Som1}
S[g_{\mu\nu},  {{X}}^a] =  \int  \bar L 
(\bar g_{\mu\nu}, \bar {{\ae}}^\mu_\nu, \si)  \sqrt{-\bar g }\, d^n x , 
\ee
with  $\bar L$ being an effective Lagrangian defining,
in the line  with ${\bar {w}}$, a particular dark gravity model.
Due to  not all of the variables in such a  form of the effective action   being independent, 
the action  is ultimately still a function of the independent basic variables 
$g_{\mu\nu}$ and $X^a$.
At last,  by means of the frame $X^a_\mu\equiv \partial_\mu X^a$  and  its inverse
$X_a^\mu$,   the DE tensor $\bar {{\ae}}^\mu_\nu$ 
may equivalently be converted into  the  scalar form as 
$ \bar {{H}}^a_b \equiv X_\mu^a  \bar {{\ae}}^\mu_\nu X_b^\nu$, with 
$\det (\bar {{\ae}}^\mu_\nu)= \det (\bar {{H}}^a_b)$, where 
\be
\bar {{H}}^a_c \eta^{cb}   \equiv \bar {{H}}^{ab} =   \bar g^{\ka\la} \partial_\ka {X}^a  \partial_\la {X}^b
\ee
is a modification of the original (quasi-)Higgs field  for gravity~\cite{Cham1}--\cite{Oda3}.
Ultimately, the effective action in the form~(\ref{Som1}), with the dependence on the DE field $\bar {{\ae}}^\mu_\nu$ 
expressed through  a  scalar potential $\bar V_{{\ae}}(\bar {{\ae}}^\mu_\nu)
=\bar V_{{H}}(\bar {{H}}^{ab})$, is  aimed at   
the spontaneous breaking of the Diff gauge symmetry/relativity, with  
producing the  dark  gravity components  ultimately due to the multiscalar.

\subsection{Weyl-scale enhanced Diff gauge symmetry}

Besides a set of fields,  any EFT is basically characterized by a pattern of its symmetries. 
For the generic multiscalar-metric dark gravity, the effective action  
(\ref{Som1}) is assumed to be generally covariant and 
gauge invariant under the  (full) Diff  symmetry  given by  a Lie derivative $D_\xi$.  
The latter depends  on a (infinitesimal) vector parameter $\xi^\mu(x)$
and may by default be constructed in terms  of  a covariant derivative $\na_\la(g_{\mu\nu})$ defined by 
the basic metric $g_{\mu\nu}$, subsequently expressed  through the effective fields as 
$g_{\mu\nu} =e^{-\bar {w}(\si)} \bar g_{\mu\nu}$.
Under  $D_\xi$, 
all the basic fields transform conventionally in accord with  their covariant tensor structure 
still irrespective of their nature.
To account for the different origin of the basic fields,
consider the  (local) Weyl scale transformations distinguishing such fields as follows:\footnote{A priori,  
the Weyl scale symmetry  is assumed to be the specifically metric one, 
with the conventional matter fields,  likewise $X^a$,  to be   inert under the Weyl rescalings, unless stated otherwise.} 
\be
{\De}_\zeta g_{\mu\nu}= \zeta g_{\mu\nu}, \quad     
{\De}_\zeta X^a=0,
\ee
where   $\zeta(x)$ is  an arbitrary (infinitesimal) scalar parameter,  so that\footnote{For definiteness,  
we restrict consideration here and in what follows by the spacetime dimension  $n=4$.} 
\bea
{\De}_\zeta  g&=& 4\zeta  g,\nn\\  
{\De}_\zeta  \Big(g_{\mu\nu}/(-g)^{1/4}\Big)&=&0,\nn\\
{\De}_\zeta {\ka}_{\mu\nu}={\De}_\zeta {\ka} & =&0.
\eea
For the effective  fields, this implies in turn 
\bea\label{Wscale}
{\De}_\zeta \si&=& 2\zeta,\nn\\
{\De}_\zeta \bar g_{\mu\nu}&=& \zeta (1+2{\bar {w}}') \bar g_{\mu\nu},\nn\\
{\De}_\zeta  \bar g&=& 
4\zeta (1+2{\bar {w}}')  \bar g,\nn\\
{\De}_\zeta \bar {{\ae}}^\mu_\nu&=& -\zeta (1+2{\bar {w}}') \bar {{\ae}}^\mu_\nu,
\eea
where  ${\bar {w}}'=d{\bar {w}}/d \si$.
Generally,  this reflects  the explicit Weyl scale violation in the multiscalar-metric dark gravity.\footnote{Note though, 
that the exclusively derivative couplings of $\si$  
still admit  the residual global Weyl scale symmetry  $g_{\mu\nu}\to e^{\zeta_0} g_{\mu\nu}$,  
with  the constant shifts  $\si\to \si+ 2\zeta_0$.}  
In an exceptional case ${\bar {w}}'=-1/2$, 
one independent component of $g_{\mu\nu}$, namely, the  determinant  $g$  
is missing from $ \bar g_{\mu\nu}$ and $\bar {{\ae}}^\mu_\nu$. 
At that,  the residual  dependence  of the effective action (\ref{Som1}) on $g$ is mediated only by $\si$.
Under~$g$ being a principle source of the CC problem, such a Weyl scale symmetry,
signifying otherwise the $g$-independence, 
is of the particular  importance for the CC screening  in 
an exceptional case of a generic dark gravity model to be exposed in what follows.

\section{Spontaneous  Diff gauge symmetry/relativity breaking}
\label{SectDGM}

\subsection{Generic dark gravity  model}

To be more particular with the (multiscalar) spontaneous breaking of the Diff gauge symmetry
(or, otherwise, relativity)  
consider a prototype  dark gravity model  merging 
tensor gravity and the associated dark gravity components,  with  the  effective Lagrangian  
in a restricted but still  rather general partitioned form 
\be\label{barL1}
\bar L =  \bar L_G (\bar g_{\mu\nu}, \bar {{\ae}}^\mu_\nu)+
\bar L_M(\bar g_{\mu\nu},\si, \phi_I) +\De \bar L,
\ee
consisting of  an extended  gravity Lagrangian $\bar L_G$,
an extended matter Lagrangian  $\bar L_M$ and a rest, $\De \bar L$.
By default,  $\bar L_G$  
may depend a priori arbitrarily  on 
the tensor fields  $\bar g_{\mu\nu}$ and $\bar {\ae}^\mu_\nu$ (and thus on  ${\ka}_{\mu\nu}$), 
while $\bar L_M$ is assumed to be  independent of $\bar {\ae}^\mu_\nu$ 
and to depend only minimally on  $\bar g_{\mu\nu}$, but a priori  
arbitrarily on  the  scalar graviton $\si$  and the  generic matter fields~$\phi_I$, $I=1,\dots$. 
The mixing part $\De \bar L$, dependent, generally,  on all the variables, 
is assumed to be negligible  in the main approximation.\footnote{In the spirit of EFT,
$\bar L_G$ is assumed to be of the order of the Plank mass squared, $M_{\rm Pl}^2$,
while $\bar L_M$ and $\bar \De L$ to be relatively suppressed in the scale of $M_{\rm Pl}$.}
More particularly, adopt for $\bar L_G$ the partitioned form:
\be
\bar L_G= \bar L_\La + \bar L_g (\bar g_{\mu\nu}) -\bar V_{{\ae}}( \bar {{\ae}}^\mu_\nu ),
\ee
where the conventional gravity part, 
$\bar L_g $, is taken for simplicity in the GR-like form for $\bar g_{\mu\nu}$: 
\be\label{LgLH}
\bar L_g= -\frac{1}{2}M_{\rm Pl}^2 R(\bar g_{\mu\nu}),
\ee
with  $M_{\rm Pl}=1/(8\pi G_N)^{1/2}$ being  the Planck mass and 
$\bar R= R(\bar g_{\mu\nu})$ the Ricci scalar curvature.\footnote{In principle,  any modification of
$\bar L_g$, like $f(\bar R)$, etc, is conceivable, too.}
This term  is  supplemented by a constant one 
\be\label{barLLa}
\bar  L_\La \equiv -M_{\rm Pl}^2  \bar \La,
\ee
with  $\bar \La$ being an effective Lagrangian CC.
At last, $\bar V_{{\ae}}$ is a scalar  potential dependent 
on  the traces  of the products of the DE tensor field $\bar {{\ae}}^\mu_\nu$:\footnote{At that,
$\det (\bar {{\ae}}^\mu_\nu)={\ka}/\bar g =e^{-2(2{\bar {w}}(\si)+\si)}$ is explicitly missed
being expressed, in principle, through  $\si$ accounted   in~$\bar L_M$.} 
\be\label{qH}
\bar V_{{\ae}}=M_{\rm Pl}^2\sum \bar v_{n_1 n_2,\dots} 
\Big(\mbox{tr}\, ({\bar {{\ae}}}^\mu_\nu)^{n_1} \mbox{tr}\, ({\bar {{\ae}}}^\mu_\nu)^{n_2}\dots \Big), 
\ee
with a priori arbitrary coefficients $\bar v_{n_1n_2\dots}$. 
The powers  $n_i=0,1,2,\dots$ are to be  
positive finite   for the potential to be analytic near  $\bar {{\ae}}^\mu_\nu=0$,
allowing in such a case a smooth limit to GR.
The potential may be normalized as $\bar V_{{\ae}}|_\star=0$ in a priori  arbitrary background point 
in the unbroken  or spontaneously broken phases,  say, at $\bar {{\ae}}_\star{}^\mu_\nu=0$ or
$\bar {{\ae}}_\star{}^\mu_\nu=\de^\mu_\nu$, respectively, 
defining, in turn, the proper normalization of $\bar \La$.
In what follows,
we restrict consideration by the simplest quartic  potential which proves to be sufficient 
in  the leading approximation (see, section~\ref{SectWTG}). 
The consistency of the theory in the higher-order approximations may  require  
a further elaboration of the potential including, in particular, the  fractional powers   
of  $\bar {{\ae}}^\mu_\nu$,  as well as  the negative ones  corresponding to  
$  (\bar {{\ae}}^\mu_\nu)^{-1}\equiv    
\bar {{\ae}}^{-1}{}^\mu_\nu= {\ka}^{-1}{}^{\mu\la}\bar g_{\la\nu} $, etc.\footnote{A priori, 
one may admit another mode of the spontaneous braking, with the potential $\bar V_{\ae}$ 
dependent instead of $\bar \ae^\mu_\nu$ on  its inverse $\bar \ae^{-1}{}^\mu_\nu$,
subject to the analyticity  requirement near $\bar \ae^{-1}{}^\mu_\nu=0$ 
and possessing by the flat  background  at the symmetric point 
$ \bar \ae^{-1}|_\star{}^\mu_\nu = \bar \ae|_\star{}^\mu_\nu=\de^\mu_\nu$.}   
Finally, the extended  matter Lagrangian $\bar L_M$ 
(incorporating, conceivably, some kinds of the particle DM in addition to the scalar-graviton dark component)  
remains a priori arbitrary, elaborating further  the purely-tensor dark gravity due to $\bar L_G$. 

\subsection{Generic field equations}

A straightforward way of  deriving the dynamical field equations (FEs) for  EFT of the multiscalar-metric dark gravity 
is to extremize  its effective action with respect to  the 
independent variations of the primary metric $ g_{\mu\nu}$ 
(or, rather, its inverse $ g^{-1\mu\nu}$)  
and the multiscalar $ {X}^a$,  expressing  afterwards the coefficients 
at the independent variations $ \de g^{-1\mu\nu}$  and   $ \de {X}^a$ 
through the effective fields $\bar g_{\mu\nu}$ 
(and its inverse  $\bar g^{\mu\nu}$), as well as  $\si$ (in particular,  ${\bar {w}}(\si)$).
By this token, the effective  variations look like
\bea\label{auxvar'}
\de \bar g^{\mu\nu}&=& e^{-{\bar {w}}} 
\de g^{-1\mu\nu}-{\bar {w}}' \bar g^{\mu\nu}\de \si,\nn\\
\frac{\de\sqrt{- \bar g}}{\sqrt{-\bar g}} &=&-\frac{1}{2}\bar g_{\ka\la}\de \bar g^{\ka\la}
=-\frac{1}{2} e^{-{\bar {w}}} \bar g_{\ka\la}\de g^{-1\ka\la}+2{\bar {w}}' \de \si,
\eea
with a prime meaning the  derivative with respect to $\si$, as well as 
\be
\de \bar {{\ae}}^\mu_\nu= \bar g^{\mu\la} \de{{\ka}}_{\la\nu}+
{\ka}_{\nu\la}\de\bar g^{\la\mu}.
\ee
In the above, one  should also  account for  the constraints (\ref{scal})  and (\ref{om})
(implying  $\si$ and ${\ka}_{\mu\nu}$ to be the effective  composite fields), 
resulting due to
\be
\frac{\de\sqrt{-g}}{\sqrt{-g}}=-\frac{1}{2} e^{-{\bar {w}}} 
\bar g_{\ka\la}\de g^{-1\ka\la}, \quad
\frac{\de\sqrt{-{{\ka}} }}{\sqrt{-{{\ka}} }}= \frac{1}{2}{\ka}^{-1\ka\la}\de {\ka}_{\ka\la},
\ee
in the respective variation of $\si$
\be\label{desi'} 
\de\si=\frac{\de\sqrt{-g}}{\sqrt{-g}}- \frac{\de\sqrt{-{\ka}}}{\sqrt{-{\ka}}}
=-\frac{1}{2}( e^{-{\bar {w}}}  
\bar  g_{\ka\la}\de {g}^{-1\ka\la} + {{\ka}}^{-1\ka\la}\de{{\ka}}_{\ka\la}),
\ee
where  
\be\label{omvar'}
\de   {{\ka}}_{\mu\nu}  = ({{X}}^a_\mu \de  {{X}}^b_\nu
+{{X}}^{a}_\nu \de {{X}}^b_\mu) \eta_{ab}, \ \ \de {X}^a_\mu\equiv \partial_\mu \de {X}^a.
\ee 
Extremizing then the effective action~(\ref{Som1}), 
given by the effective Lagrangian~(\ref{barL1})--(\ref{qH}), with respect to  the independent 
variations   $\de g^{-1\mu\nu}$ and 
$\de {\ka}_{\mu\nu}$ (the latter expressed finally through $\de {X}^a$), we  get 
the generic (under an  arbitrary ${\bar {w}}$) FEs:
\bea\label{FEgen1}
\bar R_{\mu\nu}-\frac{1}{4} \bar R \bar g_{\mu\nu}-
\frac{1}{M_{\rm Pl}^2}(\bar T_{\mu\nu}-
\frac{1}{4} \bar T \bar g_{\mu\nu})&=& 0 ,\nn\\
-\frac{1}{4}(1+2{\bar {w}}')( \bar R+4\bar\La + \frac{1}{M_{\rm Pl}^{2}}  \bar T) 
+ \frac{1}{M_{\rm Pl}^{2}}\frac{\de \bar L_{M}}{\de \si} &=&0 ,\nn\\
\bar\na_\la
\Big(\frac{1}{M_{\rm Pl}^2}\bar g^{\la\rho}
\frac{\partial \bar V_{ {\ae}}}{\partial\bar{ {\ae}}^\rho_\ka}{X}_{\ka a}  
+\frac{1}{8}( \bar R + 4 \bar\La+\frac{1}{M_{\rm Pl}^{2}}\bar T )
{X}^\la_a\Big) &=&0,\nn\\
 \frac{\de \bar L_M}{\de \phi_I}&=&0
\eea
for the three  sectors  of the multiscalar-metric dark gravity  
 ---   respectively, tensor-traceless, tensor-trace and multiscalar --- 
supplemented by FE for matter.
Here and in what follows,  $\partial/\partial$ and $\de/\de$ mean, respectively,  the partial and total (incorporating the derivatives with respect 
to the derivatives of the proper fields) variational derivatives.\footnote{At that, the scalar-graviton FE 
as such is absent, with   the composite $\si$ being, generally,  off-mass-shell, $\de \bar L_M/\de \si\neq 0$.} Besides, 
$X_{\ka a}\equiv X^b_\ka\eta_{ba}$ and   
\be\label{commonT}
\bar T_{\mu\nu}\equiv \bar T_{ {\ae}\mu\nu}+ \bar T_{M\mu\nu} ,\quad
\bar T\equiv \bar g^{\ka\la} \bar T_{\ka\la}=\bar T_{ {\ae}}+\bar T_{M},
\ee
is the total energy-momentum tensor for the DE field $\bar  {\ae}^\mu_\nu$ 
and the extended matter $M=(\si, \phi_I)$,
with the partial contributions, respectively,  as follows:
\bea\label{barTs}
\bar T_{ {\ae}\mu\nu}&\equiv& 
-\Big(\frac{\partial\bar V_{ {\ae}}}{\partial \bar  {\ae}^\mu_\la} \bar g_{\rho\nu}
+\frac{\partial \bar V_{ {\ae}}}{\partial \bar  {\ae}^\nu_\la}  \bar g_{\rho\mu}\Big)
 \bar {\ae}^\rho_\la   +\bar V_{ {\ae}}\bar g_{\mu\nu},\nn\\
\bar T_{M\mu\nu}&\equiv& \frac{2}{\sqrt{-\bar g}}
\frac{\partial(\sqrt{-\bar g}\bar L_{M})}{\partial \bar g^{\mu\nu}} =
2\frac{\partial \bar L_{M}}{\partial \bar g^{\mu\nu}}-
\bar L_{M}\bar g_{\mu\nu}.
\eea 
A priori, the derived generic FEs do not automatically imply  the 
covariant conservation of the total energy-momentum tensor, to be discussed below.

\subsection{Geometry consistency condition}

To explicitly  account for the Riemannian structure of the spacetime  
in terms of  the effective metric $\bar g_{\mu\nu}$, 
apply the proper covariant derivative $\bar\na_\mu$  to the sum of 
the  trace-free and trace (times $\bar g_{\mu\nu}$) parts of the tensor-gravity FEs.
Due to  the reduced Bianchi  identity, 
$\bar\na^\la  (\bar R_{\la\nu}-1/2\, \bar R \bar g_{\la\nu})=0$,
this gives a geometrical consistency condition
\be\label{conserve1}
-\frac{1}{2}\partial_\mu\Big({\bar {w}}'(\bar R + 
4\bar \La+ \frac{1}{ M_{\rm Pl}^2} \bar T)\Big)
= \frac{1}{M_{\rm Pl}^2}  \bar\na^\la \bar {\Theta}_{\la\mu}  ,
\ee
to be satisfied for all the solutions to the multiscalar-metric FEs.
In the above, the tensor 
\be \label{Theta1}
\bar  {\Theta}_{\mu\nu} \equiv 
\bar T_{\mu\nu} +    \bar \theta_{s\mu\nu}=
\bar T_{ {\ae}\mu\nu} + \bar T_{M\mu\nu}+   \bar \theta_{s\mu\nu}
\ee
is an extension   of the  total  energy-momentum 
tensor accounting also for the  off-mass-shell  contribution 
\be\label{theta_small}
 \bar \theta_{s\mu\nu}\equiv  -\de \bar L_M/\de \si \, \bar g_{\mu\nu}
 \ee
of the (composite) scalar graviton $\si$.\footnote{Note 
that the value $\bar \theta_{s\mu\nu}\neq 0$ 
additionally   closes or opens a spacetime region, i.e., acts in this region 
as an effective  dark matter (DM) or DE,
depending on $\de \bar L_M/\de \si$ being   positive or negative, respectively.}
A priori,
$\bar  {\Theta}_{\mu\nu}$ is not bound to covariantly conserve, 
likewise any its part~(\ref{Theta1}) separately,  in particular, $\bar \theta_{s\mu\nu}$.
The covariant conservation of $\bar  {\Theta}_{\mu\nu}$  
can still be imposed at an arbitrary $\bar{w}$ as an additional restriction on the solutions. 
The two extreme versions of the so constructed generic dark gravity model, with $\bar {w}'=0$ and $-1/2$,
presenting  the special interest --- respectively, as 
the spontaneously broken GR  vs.\ WTR ---   
are worked out in more detail below.

\subsection{Linear approximation}
\label{LAbar}

To illustrate the spontaneous breaking of the Diff gauge symmetry/relativity 
expand the basic fields $g_{\mu\nu}$ and ${X}^a$
on the background ones $g_{\star\mu\nu}$ 
and ${X}_\star^a$ supplemented by some perturbations:
\be
g_{\mu\nu}\equiv g_{\star\mu\nu}(x)+ h_{\mu\nu}(x),\quad 
{X}^a\equiv{X}_\star^a(x) +{\chi}^a(x).
\ee
Choose then the proper background coordinates $x_\star^\al\equiv \de^\al_a X^a_\star(x)$, 
with the flat background looking in such coordinates  as follows:
\be 
g_{\star\al\beta}= 
\bar g_{\star\al\beta}={\ka}_{\star\al\beta}= \eta_{\al\beta},\quad  
\bar{{\ae}}^{\pm 1}_\star{}^\al_\beta = 
\de^\al_\beta,\quad 
\si_\star=0,
\ee
where $\eta_{\al\beta}$ is the Minkowski symbol. 
Identifying the global Lorentz indices $a, b,\dots$ with the flat background ones $\al,\beta,\dots$
results then  for the basic fields  in the linear approximation (LA) in the relations:
\bea
g^{\pm1}_{\al\beta}&=&\eta_{\al\beta}\pm h_{\al\beta}({x}_\star), \nn\\ 
{X}^{\pm1}{}_\al^\beta&=& \de_\al^\beta \pm \partial_\al{\chi}^\beta({x}_\star),\nn\\ 
{\ka}^{\pm1}_{\al\beta}&=& 
\eta_{\al\beta} \pm k_{\al\beta}({x}_\star),\nn\\
 k_{\al\beta}  &\equiv&  \partial_\al{\chi}_{\beta}({x}_\star) + \partial_\beta{\chi}_{\al}({x}_\star), 
\eea
with $\partial_\al\equiv \partial/\partial x_\star^\al$ and $\chi_\al\equiv \eta_{\al\beta}\chi^\beta$, etc. 
Here and in what follows, the tensors are explicitly marked vs.\ 
their inverse values by the  superscripts $\pm 1$. 
For the effective fields, this  in turn implies in LA:
\bea\label{omf}
\bar g^{\pm1}_{\al\beta}&=& \eta_{\al\beta} \pm\bar  h_{\al\beta},  \nn\\ 
\bar{{\ae}}^{\pm1}{}^\al_\beta &=&\de^\al_\beta \mp\bar {f}^\al_\beta 
\eea
and $\si ={s}$. Designating in LA  $\bar {w} =\bar\al \si$, with $\bar \al$ a constant 
interpolating between $\bar\al=0$ for the modified GR and $\bar\al=-1/2$ for the modified WTR,  
we get in the same approximation:
\bea\label{hf'}
\bar h_{\al\beta} &\equiv & h_{\al\beta}+ 
\bar\al/2 (h- 2 \partial{\chi}) \eta_{\al\beta},       \nn\\
\bar {f}_{\al\beta} \equiv \eta_{\al\ga}\bar {f}^\ga_\beta
& =&\bar h_{\al\beta}  - k_{\al\beta}, 
\quad  \bar {f}= (1+\bar 2\bar \al)(h -2\partial\chi),
\nn\\
{s}&=& 1/2 (h-2\partial \chi),
\eea
with 
$\partial \chi\equiv \partial_\ga \chi^\ga$, 
$h=h_{\al\beta}\eta^{\al\beta}$,
$\bar f=\bar f_{\al\beta}\eta^{\al\beta}$, etc.
By construction, under  the Diff gauge symmetry enhanced by the Weyl rescalings, 
for the basic fields there takes place:
\bea\label{gaugetrans}
{\chi}^\al &\to& {\chi}^\al +\xi^\al, \nn\\ 
{k}_{\al\beta}&\to&  
{k}_{\al\beta}+( \partial_\al \xi_\beta+ \partial_\beta\xi_\al)= 
\partial_\al (\chi_\beta+\xi_\beta)+ \partial_\beta(\chi_\al+\xi_\al) ,\nn\\
h_{\al\beta}&\to& h_{\al\beta}+(\partial_\al \xi_\beta
+ \partial_\beta\xi_\al)+\zeta \eta_{\al\beta},\nn\\
h&\to& h+2\partial \xi+4\zeta,
\eea
with the arbitrary functions  $\xi^\al(x_\star)$ and $\zeta(x_\star)$ as  the  gauge parameters.  
For the effective fields, this results in LA in  transformations:
\bea\label{gauge-s}
\bar h_{\al\beta}&\to& \bar h_{\al\beta}+(\partial_\al \xi_\beta
+ \partial_\beta\xi_\al)+(1+2\bar\al)\zeta \eta_{\al\beta},\nn\\
\bar {f}_{\al\beta}&\to& \bar {f}_{\al\beta} +(1+2\bar\al)\zeta \eta_{\al\beta},\nn\\
{s}&\to& {s}+2\zeta.
\eea
Otherwise, 
$\bar h_{\al\beta}$ is a tensor under Diff, coinciding in the ``unitary'' gauge $\chi^\al=0$ (and thus $k_{\al\beta}=0$) 
according to (\ref{hf'}) with the Diff gauge-invariant $\bar {f}_{\al\beta}$,
whereas $s$ is a scalar under Diff. 
Hence,  the fields  $\bar {f}_{\al\beta}$ 
and ${s}$ may be used to  describe   
the spontaneous Diff gauge symmetry/relativity breaking in  the dark gravity in the explicitly Diff gauge-invariant terms.
At that,  $\bar f_{\al\beta}$ at $\bar\al\neq -1/2$  and $s$ explicitly violate the Weyl-scale  gauge symmetry, 
with  $s$  serving,   in effect,
as  a Weyl  scale in disguise.\footnote{To be more concise, the scalar graviton  
being a measure of  the (Weyl) scale transformations  
may be called the {\em systolon},
with the tensor graviton being conventionally  the  graviton.}

\section{Spontaneously broken General Relativity}

Let first ${\bar {w}}'=0$, so that without loss of generality ${\bar {w}}={{w}}=0$ and 
$\bar g_{\mu\nu}= g_{\mu\nu}$,
with the bar-sign to be dropped-off everywhere. 
In the Lagrangian $L=L_G+L_M$, both $L_G$ and $L_M$
are Diff gauge invariant, with the explicitly violated  Weyl scale symmetry. 
The  FEs (\ref{FEgen1}) can now be combined  as follows:
\bea\label{FEsGR}
R_{\mu\nu}-\frac{1}{2} R  g_{\mu\nu}- \La g_{\mu\nu}-
\frac{1}{M_{\rm Pl}^2} {\Theta}_{\mu\nu}  &=& 0,\nn\\
\na_\la \Big(g^{\la\rho}\frac{\partial V_{ {\ae}}}{\partial  {\ae}^\rho_\ka}{X}_{\ka a}  
+ \frac{1}{2}\frac{\de L_{M}}{\de \si}{X}^\la_a \Big)&=&0, \nn\\
 \frac{\de L_M}{\de \phi_I}&=&0,
 \eea
with  the scalar-graviton FE $\de L_{M}/\de \si=0$ as such being, generally,  missing.
Due to the reduced Bianchi identity, the extended total energy-momentum tensor 
${\Theta}_{\mu\nu}$ 
should for the geometrical consistency
be covariantly conserved for all the solutions to FEs, 
$\na^\la {\Theta}_{\la\nu}\equiv 0$, though  separately each of its  three contributions  to  the total sum~(\ref{Theta1})
may, generally,  not satisfy this requirement.
In general,  FEs~(\ref{FEsGR}) correspond  to the multiscalar-modified/spontaneously broken GR, 
with the  Lagrangian CC~$\La$
supplemented by the emergent tensor- and scalar-graviton  dark components contributing 
through ${\Theta}_{\mu\nu}$. 
Under the conventional FE  for the scalar graviton~$\si$, $ \de L_M/\de\si= 0$,
and   the absence of the tensor DE, $V_ {\ae}\equiv 0$,  
the multiscalar part of FEs~(\ref{FEsGR}) is  clearly trivial, 
with the tensor FE  reducing due to  $T_{\ae\mu\nu}=0$ to that  of the unbroken GR 
supplemented by a  scalar field $\si$. 

In LA  at $\bar \al=\al=0$,   one has $\bar h_{\al\beta}= h_{\al\beta}$, with 
the tensor and scalar dark gravity fields
\bea
{f}_{\al\beta}
& =&h_{\al\beta}  - k_{\al\beta}, 
\quad  {f}=h -2\partial\chi,
\nn\\
{s}&=& 1/2\, (h-2\partial \chi)
\eea
transforming under the enhanced Diff gauge symmetry as follows:
\bea
h_{\al\beta}&\to& h_{\al\beta}+(\partial_\al \xi_\beta
+ \partial_\beta\xi_\al)+\zeta \eta_{\al\beta},\nn\\
{f}_{\al\beta}&\to& {f}_{\al\beta} +\zeta \eta_{\al\beta},\quad 
{s}\to {s}+2\zeta,
\eea
resulting in  the Weyl scale symmetry to be explicitly violated.
At that, the  trace metric part $h$ enters both the   Diff gauge-invariant tensor field $f_{\al\beta}$ and scalar   $s$,
and, generally,  results in a ghost.
In the case with the  missing $\si$, 
$ \de L_M/\de\si\equiv 0$,   under the proper choice of the potential $V_{\ae}$,  
there appears in the broken phase   the massive tensor graviton, with  $h$ as a ghost  missing.
But the ghost inevitably  reappears 
in a sufficiently high order of the perturbation theory~\cite{Cham1}--\cite{Cham3}.
This may, conceivably, be traced back to the absence of a protecting symmetry. 
On the other hand, in the reduced case $V_{\ae}=0$  the tensor graviton remains massless, 
with  $h$ in the transverse  gauge $\partial \chi=0$ becoming 
the physical massive scalar graviton.\footnote{For the reduced case  $V_ {\ae}\equiv 0$,  
with the (ultralight) physical  $\si$ serving as a scalar DE superseding the Lagrangian CC, see~\cite{Pir3}. For the similar case,  with $\si$ as a scalar DM, cf.~\cite{Pir4}.} 
The general tensor-scalar  mixed  case beyond LA, with both $V_{\ae}\neq 0$  and $ \de L_M/\de\si\neq  0$, 
is to be investigated.

Altogether, though being able to produce the dark gravity components, 
the  spontaneously broken  GR 
at its face value can hardly help in solving the entire CC problem,
with the latter still to be resolved by some other means.
Nevertheless, an alternative to GR classically equivalent
to the latter up to CC  --- WTR --- proves  a priori  to be  well suited  
to this end, as is discussed in more detail below, with  the case of GR 
reserved as the canonical reference one. 

\section{Spontaneously broken Weyl Transverse Relativity}
\label{SectWTG}

\subsection{Full nonlinear theory}

\subsubsection{General field equations}

Let then ${\bar {w}}'=-1/2$, with the effective gravity scale factor ${\bar {w}}={\hat {w}}\equiv -\si/2$. 
Replacing in this case everywhere the bar-sign  by the hat, one has for the effective metric 
\be\label{hat}
\hat g_{\mu\nu}= e^{-\si/2} g_{\mu\nu}=({\ka}/g)^{1/4}  g_{\mu\nu},\quad \hat g=\ka,
\ee
with $\det(\hat  {\ae}^\mu_\nu) ={\ka} /\hat g=1$.
In the  effective Lagrangian $\hat L= \hat L_G +\hat L_M$, the purely tensor-gravity part $\hat L_G$ 
possesses by the Diff gauge symmetry enhanced by the ``hidden''  local Weyl scale symmetry.  
Due to this,  one  independent component in $\hat L_G$ is,  in fact, missing. 
With account for  $\hat L_M$, the missing  component 
reappears again as $\si$ signifying the explicit violation of the local Weyl scale symmetry.
The FEs (\ref{FEgen1}) now read as follows:
\bea\label{FEgenWTG1}
\hat R_{\mu\nu}-\frac{1}{4} \hat R \hat g_{\mu\nu}-
\frac{1}{M_{\rm Pl}^2}(\hat T_{\mu\nu}-\frac{1}{4} \hat T \hat g_{\mu\nu})  &=&0 ,\nn\\
\hat\na_\la
\Big(\frac{1}{M_{\rm Pl}^2}\hat g^{\la\rho}
\frac{\partial \hat V_{ {\ae}}}{\partial\hat{ {\ae}}^\rho_\ka}{X}_{\ka a}  
+\frac{1}{8}(\hat R + 4 \hat \La+\frac{1}{M_{\rm Pl}^{2}}\hat T  )
{X}^\la_a\Big) &=&0,\nn\\
\frac{\de \hat L_M}{\de \si}= \frac{\de \hat L_M}{\de \phi_I}&=&0,
\eea
with $\hat T_{\mu\nu}=\hat T_{ {\ae}\mu\nu} + \hat T_{M\mu\nu} $ given by eq.~(\ref{barTs}).
The tensor-gravity FE~(\ref{FEgenWTG1}) is trace-free, being  
principle in the context of the CC problem.\footnote{For  the viability of the trace-free tensor-gravity FE in cosmology,
cf., e.g.,~\cite{Ellis1, Ellis2}.}  
In the multiscalar-metric framework,  
such a FE alone is, in fact,  incomplete,
to be supplemented by the multiscalar~one.
At that. the Lagrangian CC $\hat \La$ clearly influences the tensor gravity
just indirectly through the multiscalar FE. 
A crucial  point here is 
that  the dynamical measure  $\sqrt{-\ka}$, becoming, in a sense, non-gravitating, 
depends  entirely on the multiscalar $X^a$,
but not on the basic metric $g_{\mu\nu}$. 
The scalar graviton  behaves now likewise a conventional scalar particle, 
with  ${ \hat \theta}_{s\mu\nu}\equiv  -\de \hat L_M/\de \si \, \hat g_{\mu\nu}=0$.
Due to the general covariance  and the conventional FEs  for $\si$ and $\phi_I$, the extended matter 
energy-momentum tensor 
$\hat T_{M\mu\nu}$ should be covariantly conserved, $\hat \na^\la\hat T_{M\la\nu}=0$,
whereas the DE $\hat T_{ {\ae}\mu\nu}$ and thus the total $\hat T_{\mu\nu}$  
may, generally,  not satisfy this requirement.\footnote{In addition to  the scalar graviton $\si$, 
the extended matter Lagrangian 
$\hat L_M= \hat L_M(\hat g_{\mu\nu}, \si, \phi_I)$ could, in principle,   account 
for the (more conventional) DM components.
In the absence of matter, the purely scalar-graviton Lagrangian
$\hat L_M=\hat L_s(\hat g_{\mu\nu}, \si)$ could be looked-for, e.g., in a most general form for the  scalar field $\si$
satisfying the  second-order FE to avoid the  explicit Ostrogradsky instabilities~\cite{Berg, Horn}.}

\subsubsection{Covariant conservation condition}

Impose additionally on a  class of the solutions to FEs~(\ref{FEgenWTG1}) 
the condition  of  the covariant conservation  of the total energy-momentum tensor 
$ \hat {\Theta}_{\mu\nu} =  \hat T_{\mu\nu} $ in  eq.~(\ref{conserve1}),
resulting upon integration in  the relation
\be
\hat R + 4\hat \La_0+ \frac{1}{ M_{\rm Pl}^2} \hat  T=0,
\ee
where $\hat \La_0$ is an integration constant, in particular, $\hat \La_0=0$ or the Lagrangian CC $\hat \La$.
Incorporating this relation refines, in turn,  the tensor-gravity and multiscalar FEs as follow:
\bea\label{CM1}
\hat R_{\mu\nu}-\frac{1}{2} \hat R  \hat g_{\mu\nu}-\hat\La_0 \hat g_{\mu\nu}- 
\frac{1}{M_{\rm Pl}^2}\hat T_{\mu\nu} &=&0, \nn\\
\hat \na_\la \Big(\frac{1}{ M_{\rm Pl}^2 }
\hat g^{\la\rho}\frac{\partial \hat V_{ {\ae}}}{\partial \hat{ {\ae}}^\rho_\ka}{X}_{\ka a}  
+\frac{1}{2}(\hat \La-\hat \La_0){X}^\la_a\Big)&=&0,\nn\\
\frac{\de \hat L_M}{\de \si}= \frac{\de \hat L_M}{\de \phi_I}&=&0,
\eea
with the Lagrangian CC $\hat \La$ clearly screened from the tensor-gravity FE, and 
$\hat T_{\mu\nu}$ 
explicitly  covariantly conserved due to the reduced Bianchi identity.
Moreover,  due to $ \hat\na^\la\hat T_{M\la\nu}=0$ there should now  separately fulfill 
$ \hat\na^\la\hat T_{ {\ae}\la\nu}=0$.
Under an arbitrary $\hat V_ {\ae}$, FEs (\ref{CM1})  describe  
the multiscalar-modified/spontaneously broken WTR, 
with DE manifesting itself through the covariantly conserved $\hat T_{ {\ae}\mu\nu}$, 
as well as the induced and Lagrangian CCs~$\hat \La_0$ and $\hat \La$, respectively. 
In the case $\hat V_ {\ae}\equiv 0$, 
the multiscalar part of FEs  (\ref{CM1})
factorizes due to $\hat g ={\ka}$ as  $\partial_\la (\sqrt{-{\ka}} X^\la_a)=0$. 
By this token, $ { \ka} $ can be chosen arbitrary, 
looking formally as non-dynamical (similarly to UR).
With account for $\hat T_{{\ae}\mu\nu}=0$, FEs  in this case reduce to  those of UR, 
equivalent classically to GR up to CC, with  the matter set (containing, conceivably, DM)
extended through the scalar graviton~$\si$ as a conventional scalar particle.\footnote{Otherwise, 
the multiscalar-modified  UR 
could be obtained as a reduction of the  multiscalar-modified WTR under imposing ab initio the constraint $g ={\ka}$, 
consistent with $\hat g ={\ka}$. This would imply, in particular, the explicit absence  of the scalar graviton $\si$, 
which is to  be added  ad hock as an additional  scalar particle.}

\subsection{Weak-field limit}
\label{LA}

\subsubsection{Linear approximation}

At $\bar\al=\hat \al=-1/2$,  in the flat background  one has in LA
\bea\label{hf}
\hat h_{\al\beta} &\equiv & h_{\al\beta}- 
1/4\, (h- 2 \partial{\chi}) \eta_{\al\beta},     
\quad \hat  h =2\partial \chi,    \nn\\
\hat {f}_{\al\beta} 
& =&\hat h_{\al\beta}  - k_{\al\beta},
\quad  \hat {f}=0,\nn\\
{s}&=& 1/2\, (h-2\partial \chi),
\eea
Under the Weyl-scale enhanced Diff gauge  transformations one gets:
\bea\label{gauge-s}
\hat h_{\al\beta}&\to& \hat h_{\al\beta}+(\partial_\al \xi_\beta
+ \partial_\beta\xi_\al),\nn\\
\hat {f}_{\al\beta}&\to& \hat {f}_{\al\beta},\quad {s}\to {s}+2\zeta.
\eea
This means  that $\hat h_{\al\beta}$, being a tensor under Diff,  is  Weyl-scale invariant, 
$\hat {f}_{\al\beta}$ is completely invariant under the Weyl-scale enhanced Diff gauge symmetry, 
while ${s}$, being a scalar under Diff, 
transforms inhomogeneously under the Weyl rescalings.
Hence, the fields  $\hat {f}_{\al\beta}$ and ${s}$  may be used to 
describe   the spontaneously  broken  Diff gauge symmetry  in  the explicitly Diff-invariant terms 
realizing the unitary gauge ($\chi^\al=0$, $\hat {h}_{\al\beta} = \hat {f}_{\al\beta}$) 
for the dark gravity.
At that, one independent DE component in the tensor $\hat {f}_{\al\beta}$
is missing due to $\hat {f}=0$ as a manifestation  of the Weyl scale symmetry.
In the end, such a  component  reappears again through the  scalar graviton~${\si}$ as a conventional particle, 
reflecting ultimately the explicit local Weyl scale violation.\footnote{Note that imposing 
ab initio the UR restriction $g={\ka}$ eliminating, in fact, one independent component 
would result  in LA in $h =\hat h=2\partial \chi$ (still preserving $\hat {f}=0$) 
and $\hat h_{\al\beta}= h_{\al\beta}$, as well as 
${s}=0$, with the lost of the Weyl scale symmetry.}
The   account beyond LA for the higher orders 
in the presence of the protecting Weyl scale symmetry,
as well as the local Weyl-scale violation due to $\si$, are to be investigated.
  
\subsubsection{Massive tensor graviton}

To clarify the viability of the multiscalar-modified/spontaneously broken WTR as a theory of gravitons 
retain first only the Weyl-scale  invariant purely tensor-gravity Lagrangian~$\hat L_G$
putting $\hat L_M(\si,\phi_I)=0$ in neglect by  matter and the explicit scalar-graviton  contribution. 
Choosing then as independent variables for the DE tensor field ${\hat  {\ae}}^\mu_\nu$  its trace $\hat  {\ae}$ 
and the traceless part $ \hat {\tilde  {\ae}}^\mu_\nu$, respectively, as
\be
\hat  {\ae}\equiv  {\hat  {\ae}}^\ka_\la\de_\ka^\la,  \quad 
\hat{\tilde   {\ae}}^\mu_\nu \equiv \hat  {\ae}^\mu_\nu-\frac{1}{4}\hat  {\ae}\de^\mu_\nu , 
\ee
consider  the even quartic potential $\hat V_{\ae}( \hat{\tilde   {\ae}}^\mu_\nu, \hat  {\ae}) $  
as follows:
\be\label{quadpot}
\hat V_ {\ae}=
\frac{1}{2}M_{\rm Pl}^2 \hat v_2  \bigg( \hat{\tilde   {\ae}}^\mu_\nu \hat{\tilde   {\ae}}^\nu_\mu 
 +\hat\beta_4 \Big((\frac{1}{4} \hat  {\ae})^2-1\Big)^2  \bigg),   
\ee
where  $\hat v_2>0$ and $\hat \beta_4$ are some constant parameters.\footnote{In particular,
a similar potential $V_{\ae} $ at a proper $\beta_4$ 
is used for the consistent description of the massive tensor graviton within  the spontaneously broken GR~\cite{Cham1}.}
For definiteness, $\hat V_{\ae}$  is normalized  as $\hat V_{\ae}|_\star =0$  in the flat  background at 
$\hat  {\ae}_\star{}^\mu_\nu = \de^\mu_\nu$, $\hat  {\ae}_\star=4$,  with 
an additional constant contribution  assigned by default  
to the Lagrangian CC~$\hat \La$. 
Such a  potential   is peculiar by possessing  by  the two extremums under $\hat \ae$: at $\hat\ae=0$ and $|\hat \ae|=4$,
which may be ascribed at $\hat \beta_4>0$ to the unbroken trivial and broken flat backgrounds, respectively.
In the trivial background at $\hat  {\ae}_\star{}^\mu_\nu=0$,  $\hat {\ae}_\star=0 $, the potential $\hat V_{\ae}|_\star$
at  $\hat \beta_4>0$  lies higher, being positive 
and unstable near $\hat {\ae}_\star=0 $.
In a vicinity of the flat background at $\hat  {\ae}_\star{}^\al_\beta=\de^\al_\beta$, 
the potential $\hat V_{\ae}$
in the quadratic  approximation is as follows:\footnote{In fact, due to $\hat {f}=0$ in LA
the second term in~(\ref{h2})  in the quadratic  approximation is  irrelevant, 
with the potential being (quasi-)stable  irrespective of $\hat\beta_4$, 
admitting, in particular,  $\hat \beta_4=0$ and the purely-quadratic $\hat V_{\ae}(\hat{\tilde  {\ae}}^\mu_\nu)$.}
\be\label{h2}
 \hat V_ {\ae}=\frac{1}{2} M_{\rm Pl}^2\hat v_2 \hat {f}^\al_\beta \hat {f}_\al^\beta 
+\frac{1}{4} (\hat \beta_4-1)\hat {f}^2 ,
\ee
being  in LA due to $\hat f=0$,  in fact, independent of $\hat f$ irrespective of $\hat\beta_4$. 
This gives in turn: 
\be
 \hat T_{ {\ae}\al\beta}= 2 M_{\rm Pl}^2\hat v_2 \hat {f}_{\al\beta},\quad 
 \hat T_{ {\ae}}\equiv \hat T_{ {\ae}\al\beta}\eta^{\al\beta}=0.
\ee
Due to the Diff gauge symmetry, accounting  
under the unitary  gauge $\chi^\al = 0$ (and thus $k_{\al\beta}=0$) for (\ref{hf}), 
one can substitute in $\hat R_{\al\beta}$ 
the tensor-graviton field $\hat h_{\al\beta}$,  with $\hat h=0$,
by the DE  one $\hat {f}_{\al\beta}$,  with $\hat {f}=0$, getting in~LA
\be
\hat R_{\al\beta} =
\frac{1}{2}(\partial_\al\partial_\ga \hat {f}^\ga_\beta+  
\partial_\beta\partial_\ga \hat {f}^\ga_\al
-\partial^2\hat {f}_{\al\beta}),\quad  \hat R=\partial_\ga\partial_\de \hat {f}^{\ga\de}.
\ee
This results in LA in the tensor and multiscalar FEs (\ref{FEgenWTG1}) as follows:
\bea\label{linear'}
\partial_\al\partial_\ga \hat {f}^\ga_\beta+  \partial_\beta\partial_\ga \hat {f}^\ga_\al     
-\partial^2 \hat {f}_{\al\beta}
-\frac{1}{2}\partial_\ga\partial_\de \hat {f}^{\ga\de}\eta_{\al\beta}&=
&m_g^2\hat {f}_{\al\beta},\nn\\
\partial_\al (m_g^2 \hat {f}^\al_\beta  -
\frac{1}{2}\partial_\ga\partial_\de \hat {f}^{\ga\de}\de^\al_\beta)&=&0,
\eea
clearly independent of $\hat \La$. The latter property 
can be explicitly confirmed under an arbitrary $\chi^\al$ due to the Diff gauge symmetry.  
In the above,  we have put 
$\hat v_2\equiv  m_g^2/4$, with $m_g$ to be associated with the tensor-graviton mass. 
It proves that the multiscalar FE follows, in fact, from the tensor one, 
with all the nine dark-gravity  components remaining thus independent. 
At that, one  missing
component corresponds to the neglected scalar graviton, the 
inclusion of which would explicitly violate  the hidden Weyl scale symmetry.
To reduce the number of the independent components, 
impose additionally the covariant conservation condition (\ref{conserve1}) 
resulting in LA in the transversality condition
\be\label{df0}
\partial_\ga\hat {f}^{\ga\al}=0
\ee
followed by   $\hat \La_0=0$.
This finally results in  FE 
\be\label{d2f}
(\partial^2 +m_g^2)\hat {f}_{\al\beta}=0
\ee
for the gauge-invariant $\hat  {f}_{\al\beta}$ coinciding with the gauge-variant 
$\hat  h_{\al\beta}=   h_{\al\beta}-1/4\, h \eta_{\al\beta} $ under the 
unitary gauge $\chi_\al=0$.
The FEs   (\ref{df0}) and (\ref{d2f}) describe precisely the five independent DE 
degrees of freedom without ghosts, similarly to  the Fierz-Pauli case 
for the massive  tensor graviton within the spontaneously broken GR~\cite{Cham1}.
The covariant conservation condition 
in the  spontaneously broken  WTR, likewise GR,  proves  thus to be of principle  
to ensure the consistency of the massive tensor graviton description.
Still, the conceivable violation of the covariant conservation condition 
in the context of DE is of particular interest.

\subsubsection{Massless limit}

At last, in the massless limit $m_g\to 0$, in  FEs~(\ref{linear'}) there appears 
a three-parameter residual gauge  symmetry
\be
\hat {f}_{\al\beta}\to \hat {f}_{\al\beta}-(\partial_\al \hat \varphi_\beta+ 
\partial_\beta \hat \varphi_\al),  \quad \partial\hat \varphi=0, \quad 
 \partial^2\hat \varphi_\al=0,  
\ee
satisfied also by the constraint~(\ref{df0}). This  reduces the number of the independent 
tensor-graviton components on-mass-shell $m_g=0$ 
to two,  making thus the massless limit in LA consistent.
In fact, in this limit $\hat {f}_{\al\beta}$ reduces to the  conventional massless 
unimodular  $ \hat h_{\al\beta}$, with $\hat h=0$.
The presence of the 
residual gauge symmetry ensuring in the spontaneously broken WTR 
the smooth massless  limit  to UR
is thus  crucial  for the explicit consistency in LA of the massive tensor-graviton description.

\subsubsection{Scalar graviton}

Finally,  retaining the  tensor-graviton description 
through the Weyl-scale invariant Lagrangian $\hat L_G$, with the three (at $n=4$)  gauge components 
converted into the additional physical ones for the  massive tensor graviton,
one could add the explicitly Weyl-scale violating $\hat L_M(\si)$ to consistently realize  the last  component 
contained in the multiscalar~$X^a$.
In the neglect by matter,  
this would result  in LA in FE for the scalar graviton~$s$ 
as a conventional particle:
\be
(\partial^2 +m_s^2) s=0, 
\ee
with  a scalar-graviton mass $m_s$. In the above, 
the gauge-invariant scalar field $s$ coincides with the gauge-variant metric component 
$h/2 $ under the transverse gauge $\partial \chi=0$ (embodying, in particular,  the unitary one $\chi^\al=0$).
Going beyond LA and accounting for the mixing of the scalar graviton $\si$ with matter remain to be studied.

\section{Conclusion}

The generic multiscalar-metric framework provides 
a promising  route for the  modification of  gravity, with 
the spontaneous breaking of the  Diff gauge symmetry/relativity 
as a basic reason  of the dark gravity -- a~merger 
of gravity with  the emergent dark components.   
The multiscalar-modified/spontaneously broken 
WTR, built on  this route,  incorporating the Weyl scale symmetry
and the Higgs mechanism for gravity,   
may well serve as a viable beyond-GR  EFT of gravity. It is able  to
screen the Lagrangian CC $\hat \La$, with  emergence of the induced one $\hat \La_0$
supplemented by the massive tensor and scalar gravitons  as the dark gravity components. 
As EFT, the so constructed dark gravity model,  
being applied to the  (extensions of the) particle SM, allows, in principle, to account also
for the spontaneous breaking  of the internal symmetries. This should be followed 
by the step-wise modifications of  the Lagrangian CC 
and matching of  the emergent descriptions at the phase transition scales.
In the leading approximation, the spontaneously broken WTR  
proves  to be well-behaving as the  theory of the massive tensor and scalar gravitons without ghosts.
Further developing the basic concept of  the spontaneously broken Diff gauge symmetry/relativity 
--- in particular,  WTR vs.\ GR --- 
as a principle source of the emergent dark gravity components of the Universe  is urgent.

\paragraph{Acknowledgment} The author is  sincerely grateful to S.~S.\ Gershtein 
for encouraging discussions.

\end{document}